\documentclass[preprint,preprintnumbers, prd, floatfix, superscriptaddress,nofootinbib] {revtex4-1}
\usepackage{epsfig}
\usepackage{subfigure}
\usepackage{dcolumn}
\usepackage{bm}
\usepackage[usenames ,dvipsnames]{xcolor}
\usepackage{slashed}
\usepackage{graphicx,color}
\begin{document}
\title{Study of two-body doubly charmful baryonic $B$ decays\\
with $SU(3)$ flavor symmetry}

\author{Yu-Kuo Hsiao}
\email{Email address: yukuohsiao@gmail.com}
\affiliation{School of Physics and Information Engineering, 
Shanxi Normal University, Taiyuan 030031, China}

\date{\today}

\begin{abstract}
Within the framework of $SU(3)$ flavor symmetry, 
we investigate two-body doubly charmful baryonic $B\to{\bf B}_c\bar{\bf B}'_c$ decays,
where ${\bf B}_c\bar{\bf B}'_c$ represents the anti-triplet charmed dibaryon. 
We determine the $SU(3)_f$ amplitudes and calculate
${\cal B}(B^-\to \Xi_c^0\bar \Xi_c^-)=(3.4^{+1.0}_{-0.9})\times 10^{-5}$ and
${\cal B}(\bar B^0_s\to \Lambda_c^+\bar \Xi_c^-)=(3.9^{+1.2}_{-1.0})\times 10^{-5}$
induced by the single $W$-emission configuration.  We find that the $W$-exchange amplitude, 
previously neglected in studies, needs to be taken into account.
It can cause a destructive interfering effect with the $W$-emission amplitude, 
alleviating the significant discrepancy between the theoretical estimation and experimental data
for ${\cal B}(\bar B^0\to\Lambda_c^+\bar\Lambda_c^-)$.
To test other interfering decay channels, we calculate
${\cal B}(\bar B^0_s\to \Xi_c^{0(+)}\bar \Xi_c^{0(+)})=(3.0^{+1.4}_{-1.1})\times 10^{-4}$ and
${\cal B}(\bar B^0\to \Xi_c^0\bar \Xi_c^0)=(1.5^{+0.7}_{-0.6})\times 10^{-5}$.
We estimate non-zero branching fractions for
the pure $W$-exchange decay channels, specifically
${\cal B}(\bar B^0_s\to \Lambda_c^+\bar \Lambda_c^-)=(8.1^{+1.7}_{-1.5})\times 10^{-5}$ and
${\cal B}(\bar B^0\to \Xi_c^+\bar \Xi_c^-)=(3.0\pm 0.6)\times 10^{-6}$. Additionally, we predict
${\cal B}(B^+_c\to \Xi_c^+\bar \Xi_c^0)=(2.8^{+0.9}_{-0.7})\times 10^{-4}$ and
${\cal B}(B^+_c\to \Lambda_c^+\bar \Xi_c^0)=(1.6^{+0.5}_{-0.4})\times 10^{-5}$,
which are accessible to experimental facilities such as LHCb.
\end{abstract}

\maketitle
\section{introduction}
The tree-level dominated two-body charmless baryonic $B$ meson decays,
$B\to {\bf B\bar B'}$, can proceed through the $W$-boson exchange ($W_{\rm ex}$), 
$W$-boson annihilation ($W_{\rm an}$), and $W$-boson emission ($W_{\rm em}$) 
decay configurations. In analogy with leptonic $B$ decay, where the $W_{\rm an}$ amplitude
${\cal M}_{\rm wan}(B\to\ell\bar \nu)\propto m_\ell \bar u_\ell(1+\gamma_5)v_{\bar \nu}$
involves a tiny lepton mass $m_\ell$ corresponding to helicity suppression~\cite{Bevan:2014iga,Hou:2019uxa}, 
${\cal M}_{\rm wex(wan)}(B\to {\bf B\bar B'})\propto 
m_-\langle {\bf B\bar B'}|\bar q q'|0\rangle+ m_+ \langle {\bf B\bar B'}|\bar q\gamma_5 q'|0\rangle$
with $m_{\mp}=m_q\mp m_{q'}$ is considered to be
more suppressed than ${\cal M}_{\rm wem}(B\to {\bf B\bar B'})$~\cite{Chua:2022wmr}. 
Hence, it raises the debate if one can really neglect the $W_{\rm ex(an)}$ contribution to 
the branching fractions~\cite{Chua:2022wmr,Chernyak:1990ag,Ball:1990fw,
Chang:2001jt,Cheng:2001tr,Chua:2013zga,Hsiao:2014zza,Huang:2021qld}.

In the study of singly charmful baryonic $B\to{\bf B}_c \bar{\bf B}'$ decays,
the $W_{\rm ex(an)}$ amplitude was also neglected~\cite{He:2006vz,Cheng:2002sa}.
Nonetheless, it has been found that ${\cal M}_{\rm wex(wan)}(B\to {\bf B}_c \bar {\bf B}')\propto
m_c \langle {\bf B}_c\bar {\bf B}'|\bar c(1+\gamma_5)q|0\rangle$ with $m_c\gg m_q$
can alleviate the helicity suppression~\cite{Hsiao:2019wyd}. This results in 
${\cal B}_{\rm wex}(\bar B_s^0\to\Lambda^+_c\bar p)$ and 
${\cal B}_{\rm wex}(\bar B^0\to\Xi_c^+\bar\Sigma^-)$ being predicted to be of order $10^{-5}$,
much more reachable than ${\cal B}(B\to{\bf B}\bar{\bf B}')\sim 10^{-8}-10^{-7}$
for the test of a non-negligible $W_{\rm ex(an)}$ contribution. However, 
until very recently, these observations have not been reported.
 
It is worth noting that two-body doubly charmful baryonic $B$ decays, $B\to{\bf B}_c\bar{\bf B}'_c$,
have provided a possible experimental indication of a non-negligible contribution
from the $W_{\rm ex}$ term. The measured branching fractions for $B\to {\bf B}_c \bar{\bf B}'_c$
are reported as follows:
\begin{eqnarray}\label{data1}
{\cal B}(\bar B^0\to \Xi_c^+\bar \Lambda_c^-)
&=&(1.2\pm 0.8)\times 10^{-3}~\text{\cite{pdg}}\,,
\nonumber\\
{\cal B}(B^-\to \Xi_c^0\bar \Lambda_c^-)
&=&(9.5\pm 2.3)\times 10^{-4}~\text{\cite{pdg}}\,,
\nonumber\\
{\cal B}(\bar B^0\to \Lambda_c^+\bar \Lambda_c^-)
&<&1.6\times 10^{-5}~\text{\cite{pdg,LHCb:2014scu}}\;
\nonumber\\
&=&(2.2^{+2.2}_{-1.6}\pm 1.3)\times 10^{-5}~\text{\cite{Belle:2007lyc}}\,,
\nonumber\\
{\cal B}(\bar B^0_s\to \Lambda_c^+\bar \Lambda_c^-)
&<&9.9\times 10^{-5}~\text{\cite{pdg,LHCb:2014scu}}\,.
\end{eqnarray}
Initially, it was considered that $B\to{\bf B}_c\bar{\bf B}'_c$ receives 
a single contribution from the $W_{\rm em}$ topology~\cite{Cheng:2009yz,Belle:2007lyc}. 
In Eq.~(\ref{data1}), ${\cal B}(\bar B^0\to \Xi_c^+\bar \Lambda_c^-)
\simeq{\cal B}(B^-\to \Xi_c^0\bar \Lambda_c^-)$ seemingly 
supporting this assumption. Nonetheless, it also leads to an estimation of 
${\cal B}(\bar B^0\to \Lambda_c^+\bar \Lambda_c^-)
\simeq (V_{cd}/V_{cs})^2 (\tau_{\bar B^0}/\tau_{B^-}) 
{\cal B}(B^-\to \Xi_c^0\bar \Lambda_c^-)=(4.7\pm 1.1)\times 10^{-5}$ by utilizing
the ${\cal B}(B^-\to \Xi_c^0\bar \Lambda_c^-)$ value from Eq.~(\ref{data1}).
This clearly shows a significant deviation from the experimental upper limit of $1.6\times 10^{-5}$ 
by around 3 standard deviations. Therefore, it is reasonable to infer that
the $W_{\rm ex}$ topology, overlooked in $\bar B^0\to \Lambda_c^+\bar \Lambda_c^-$,
should be taken into account. It can cause a destructive interfering effect, thus
reducing the overestimated branching fraction. Additionally, the $W_{\rm ex}$ topology 
can induce a non-zero ${\cal B}(\bar B^0_s\to \Lambda_c^+\bar \Lambda_c^-)$,
warranting further examination.

For clarification, a careful study of $B\to{\bf B}_c\bar {\bf B}'_c$ is necessary.
The $SU(3)$ flavor symmetry can be a useful theoretical tool~\cite{Chua:2022wmr,Chua:2013zga,
Savage:1989jx,Savage:1989ub,Kohara:1990gu,He:2000ys,Fu:2003fy,
Hsiao:2015iiu,Zhao:2018mov,Hsiao:2019yur,Pan:2020qqo,Hsiao:2021nsc,
Savage:1989qr,Xing:2023dni,Huang:2021aqu,Zhong:2022exp}, allowing us to
parameterize the amplitudes without involving the complexity of model calculations.
Hence, we propose using the $SU(3)_f$ approach to specifically explore 
the $W_{\rm ex}$ and $W_{\rm em}$ contributions to $B\to{\bf B}_c\bar{\bf B}'_c$.
The first observation of $B_c^+\to J/\Psi p\bar p\pi^+$ by LHCb~\cite{LHCb:2014ebd} indicates 
a potential test-bed for the baryonic phenomena in $B_c^+$ decays,
such as the branching fraction~\cite{Hou:2000bz,Chua:2001vh,Chua:2002wn,Chua:2002yd,
Geng:2005fh,Geng:2007cw,Hsiao:2022tfj,Hsiao:2017nga,Hsiao:2022uzx},
direct $CP$ asymmetry~\cite{Geng:2006jt,Hsiao:2019ann},
triple product asymmetry~\cite{Geng:2005wt}, 
angular distribution~\cite{Huang:2022oli}, 
and exotic states~\cite{Hsiao:2015nna,Hsiao:2013dta,Hsiao:2014tda}
as studied in baryonic $B$ decays. Therefore,
we will estimate ${\cal B}(B_c^+\to {\bf B}_c\bar{\bf B}'_c)$
to initiate a theoretical investigation.

\section{Formalism}
%
\begin{figure}[t!]
\centering
\includegraphics[width=2.9in]{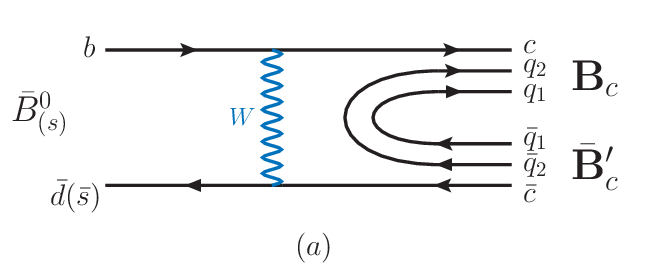}
\includegraphics[width=3in]{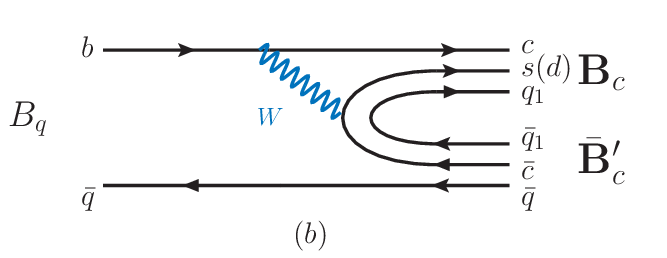}
\includegraphics[width=3in]{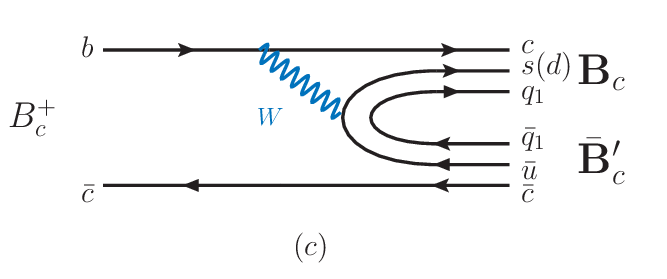}
\caption{Feynman diagrams of $B_q\to{\bf B}_c{\bf \bar B}'_c$ and $B_c^+\to{\bf B}_c{\bf \bar B}'_c$ decays.}\label{fig1}
\end{figure}
%
To study the two-body doubly charmful baryonic $B_{(c)}\to {\bf B}_c\bar {\bf B}'_c$ decays
with $B_c$ denoting $B_c^+(b\bar c)$,
the quark-level effective Hamiltonians for the $b\to c\bar q q'$ weak transitions are required,
given by~\cite{Buchalla:1995vs,Buras:1998raa}
\begin{eqnarray}\label{Heff}
{\cal H}_{eff}^{b\to c\bar q q'}&=&\frac{G_F}{\sqrt 2} V_{cb}V_{qq'}^*
\bigg[c_1(\bar q' q)(\bar c b)+c_2(\bar q'_\beta q_\alpha)(\bar c_\alpha b_\beta)\bigg]\,,
\end{eqnarray}
where $G_F$ is the Fermi constant, $V_{cb}$ and $V_{qq'}$ 
with $q=(u,c)$ and $q'=(s,d)$ the Cabibbo-Kobayashi-Maskawa (CKM) matrix elements. 
In Eq.~(\ref{Heff}), we define
$(\bar q_1 q_2)=\bar q_1\gamma_\mu(1-\gamma_5)q_2$, and
the subscripts $(\alpha,\beta)$ denote the color indices; moreover,
$c_{1,2}$ are the scale ($\mu$)-dependent Wilson coefficients with $\mu=m_b$ for the $b$ decays. 
In the $SU(3)_f$ representation, 
${\cal H}_{eff}^{b\to c\bar c q'}$ and ${\cal H}_{eff}^{b\to c\bar u q'}$ by omitting Lorentz structure 
are reduced as $H^i$ and $H^i_j$, respectively, where $i$ and $j$ run from 1 to 3
to represent the flavor indices. Explicitly, the nonzero entries are given by~\cite{Pan:2020qqo}
\begin{eqnarray}\label{Hijk}
H^2_1=\lambda_{ud}\,,\;H^3_1=\lambda_{us}\,,\;
H^2 = \lambda_{cd}\,,\;H^3 =\lambda_{cs}\,,
\end{eqnarray}
with $\lambda_{qq'}\equiv V_{cb}V^*_{qq'}$.  
Accordingly, we present the $B$ meson and ${\bf B}_c$ baryon in the $SU(3)_f$ forms:
\begin{eqnarray}\label{BqBc}
B(B_i)&=&(B^-,\bar B^0,\bar B^0_s)\,,
\nonumber\\
{\bf B}_c({\bf B}_{c}^{ij})&=&\left(\begin{array}{ccc}
0& \Lambda_c^+ & \Xi_c^+\\
-\Lambda_c^+&0&\Xi_c^0 \\
-\Xi_c^+&-\Xi_c^0&0
\end{array}\right)\,,
\end{eqnarray}
whereas $B_c$ is a singlet. 
By connecting the flavor indices of the initial state to those of the effective Hamiltonian and final states,
the $SU(3)_f$ approach yields the amplitudes to be
\begin{eqnarray}\label{amp}
{\cal M}(B\to{\bf B}_c\bar {\bf B}'_c)&=&
e B_i H^i {\bf B}_{c\;jk} {\bf \bar B}^{\prime\;jk}_{c}+
c' B_i H^j {\bf B}_{c\;jk} {\bf \bar B}^{\prime\;ik}_{c}\;,\nonumber\\
{\cal M}(B_c\to{\bf B}_c\bar {\bf B}'_c)&=&
\bar c' H^i_j {\bf B}_{c\;ik}{\bf \bar B}^{\prime\;jk}_{c}\;,
\end{eqnarray}
where the parameter $e$ and $c'(\bar c')$ correspond to 
the $W_{\rm ex}$ and $W_{\rm em}$ configurations 
in Fig.~\ref{fig1}a and Fig.~\ref{fig1}b(c), respectively.
For a later numerical analysis, we use the equation~\cite{pdg}:
\begin{eqnarray}\label{p_space}
&&{\cal B}(B_{(c)}\to{\bf B}_c\bar {\bf B}'_c)=
\frac{G_F^2|\vec{p}_{\rm cm}|\tau_{B_{(c)}}}{16\pi m_{B_{(c)}}^2 }
|{\cal M}(B_{(c)}\to{\bf B}_c\bar {\bf B}'_c)|^2\,,\nonumber\\
&&|\vec{p}_{cm}|=\frac{
\sqrt{(m_{B_{(c)}}^2-M_+^2)(m_{B_{(c)}}^2-M_-^2)}}{2 m_{B_{(c)}}}\,,
\end{eqnarray}
to compute the branching fractions,
where $M_\pm\equiv m_{{\bf B}_c}\pm m_{\bar{\bf B}'_c}$, 
$\vec{p}_{cm}$ is the three-momentum of the ${\bf B}_c$ baryon in the $B_{(c)}$ meson rest frame,
and $\tau_{B_{(c)}}$ stands for the $B_{(c)}$ lifetime. The amplitude 
${\cal M}(B_{(c)}\to{\bf B}_c\bar {\bf B}'_c)$ can be found in Table~\ref{tab1}.

%
\begin{table}[b]
\caption{Amplitudes of $B_{(c)}\to{\bf B}_c\bar {\bf B}'_c$
with the $SU(3)_f$ parameters $e$ and $c'(\bar c')$.}\label{tab1}
\footnotesize
\begin{tabular}{|l|l|}
\hline
Decay modes& Amplitudes \\
\hline\hline
$\bar B^0\to \Xi_c^+\bar \Lambda_c^-$
&$-\lambda_{cs} c'$ \\
$B^-\to \Xi_c^0\bar \Lambda_c^-$
&$\lambda_{cs} c'$ \\
$\bar B^0_s\to \Xi_c^0\bar \Xi_c^0$
&$-\lambda_{cs} (2e+c')$ \\
$\bar B^0_s\to \Xi_c^+\bar \Xi_c^-$
&$-\lambda_{cs} (2e+c')$ \\
$\bar B^0_s\to \Lambda_c^+\bar \Lambda_c^-$
&$-\lambda_{cs} (2e)$ \\
$B^+_c\to \Xi_c^+\bar \Xi_c^0$
&$-\lambda_{ud} \bar c'$ \\
\hline
\end{tabular}
\begin{tabular}{|l|l|}
\hline
Decay modes& Amplitudes \\
\hline\hline
$\bar B^0\to \Xi_c^0\bar \Xi_c^0$
&$-\lambda_{cd} (2e+c')$ \\
$\bar B^0\to \Xi_c^+\bar \Xi_c^-$
&$-\lambda_{cd}  (2e)$ \\
$\bar B^0\to \Lambda_c^+\bar \Lambda_c^-$
&$-\lambda_{cd}  (2e+c')$ \\
$B^-\to \Xi_c^0\bar \Xi_c^-$
&$-\lambda_{cd}  c'$ \\
$\bar B^0_s\to \Lambda_c^+\bar \Xi_c^-$
&$-\lambda_{cd}  c'$ \\
$B^+_c\to \Lambda_c^+\bar \Xi_c^0$
&$\lambda_{us} \bar c'$ \\
\hline
\end{tabular}

\end{table}
%

\section{Numerical Results}
%
\begin{table}[b]
\caption{Branching fractions of $B_{(c)}\to{\bf B}_c \bar{\bf B}'_c$ decays.
}\label{pre}
{
\footnotesize
\begin{tabular}{lcc}
\hline
decay channel&this work&experimental data\\
\hline\hline
$10^4 {\cal B}(\bar B^0\to \Xi_c^+\bar \Lambda_c^-)$
&$7.2^{+2.1}_{-1.9}$ 
&$12\pm 8$~\text{\cite{pdg}}\\
$10^4 {\cal B}(B^-\to \Xi_c^0\bar \Lambda_c^-)$
&$7.8^{+2.3}_{-2.0}$ 
&$9.5\pm 2.3$~\text{\cite{pdg}}\\
$10^4 {\cal B}(\bar B^0_s\to \Xi_c^0\bar \Xi_c^0)$
&$3.0^{+1.4}_{-1.1}$
&\\
$10^4 {\cal B}(\bar B^0_s\to \Xi_c^+\bar \Xi_c^-)$
&$3.0^{+1.4}_{-1.1}$
&\\
$10^5 {\cal B}(\bar B^0_s\to \Lambda_c^+\bar \Lambda_c^-)$
&$8.1^{+1.7}_{-1.5}$ 
&$<9.9$~\text{\cite{pdg,LHCb:2014scu}}\\
$10^4 {\cal B}(B^+_c\to \Xi_c^+\bar \Xi_c^0)$
&$2.8^{+0.9}_{-0.7}$
&\\
\hline
$10^5 {\cal B}(\bar B^0\to \Xi_c^0\bar \Xi_c^0)$
&$1.5^{+0.7}_{-0.6}$
&\\
$10^6 {\cal B}(\bar B^0\to \Xi_c^+\bar \Xi_c^-)$
&$3.0\pm 0.6$
&\\
$10^5 {\cal B}(\bar B^0\to \Lambda_c^+\bar \Lambda_c^-)$
&$2.1^{+1.0}_{-0.8}$ 
&$<1.6$~\text{\cite{pdg,LHCb:2014scu}} $(2.2^{+2.6}_{-2.1}~\text{\cite{Belle:2007lyc}})$\\
$10^5 {\cal B}(B^-\to \Xi_c^0\bar \Xi_c^-)$
&$3.4^{+1.0}_{-0.9}$
&\\
$10^5 {\cal B}(\bar B^0_s\to \Lambda_c^+\bar \Xi_c^-)$
&$3.9^{+1.2}_{-1.0}$
&\\
$10^5 {\cal B}(B^+_c\to \Lambda_c^+\bar \Xi_c^0)$
&$1.6^{+0.5}_{-0.4}$
&\\
\hline
\end{tabular}}
\end{table}
%
In the numerical analysis,
the CKM matrix elements are adopted from PDG~\cite{pdg}:
\begin{eqnarray}\label{ckm}
(V_{cb},V_{cs},V_{ud},V_{us},V_{cd})=(A\lambda^2,1-\lambda^2/2,1-\lambda^2/2,\lambda,-\lambda)\,,
\end{eqnarray}
where $A=0.826$ and $\lambda=0.225$ in the Wolfenstein parameterization. 
In Eq.~(\ref{amp}), the parameters $e$ and $c'$ are complex numbers,
which we present as
\begin{eqnarray}\label{para1}
|c'|, |e|e^{i\delta_e}, 
\end{eqnarray}
with $\delta_e$ a relative phase. By using the experimental data in Table~\ref{pre},
we solve the parameters as
\begin{eqnarray}\label{para2}
|c'|=(1.29\pm 0.18)~{\rm GeV}^3\,,\;
|e|=(0.20\pm 0.02)~{\rm GeV}^3\,,\; 
\delta_e=180^\circ\,.
\end{eqnarray}
Explicitly, we use the experimental results for ${\cal B}(\bar B^0\to \Xi_c^+\bar \Lambda_c^-)$ and 
${\cal B}(B^-\to \Xi_c^0\bar \Lambda_c^-)$ to fit $|c'|$. On the other hand,
the experimental data have not been sufficient and accurate enough 
to simultaneously determine $|e|$ and the relative phase~$\delta_e$, 
as indicted in Table~\ref{pre}. 
For a practical determination,
we fix $\delta_e=180^\circ$ to cause a maximum destructive interference.
As a consequence, the experimental upper bounds of 
${\cal B}(\bar B^0_s\to \Lambda_c^+\bar \Lambda_c^-)$ and
${\cal B}(\bar B^0\to \Lambda_c^+\bar \Lambda_c^-)$
can sandwich an allowed range for $|e|$, as given in Eq.~(\ref{para2}).
Moreover, we assume $\bar c'= c'$ 
due to the similarity of the Feynman diagrams in Figs.~\ref{fig1}b and \ref{fig1}c.
Subsequently, we calculate the branching fractions as provided in Table~\ref{pre}
using the determination in Eq.~(\ref{para2}).

\section{Discussion and Conclusions}
The $SU(3)_f$ approach enables us to explore all possible $B\to{\bf B}_c\bar{\bf B}'_c$ decays, 
as summarized in Table~\ref{tab1}. Furthermore, it helps in deriving constraints on $SU(3)_f$ relations, 
facilitating the decomposition of amplitudes into $e$ and $c'$ terms. 
These terms parameterize the $W_{\rm ex}$ and $W_{\rm em}$ topologies 
depicted in Fig.\ref{fig1}a and Fig.\ref{fig1}b(c), respectively.

In $b\to c\bar c s$ induced decays, the $SU(3)_f$ symmetry unequivocally establishes that 
both $\bar B^0\to \Xi_c^+\bar \Lambda_c^-$ and $B^-\to \Xi_c^0\bar \Lambda_c^-$ 
solely proceed through the $W_{\rm em}$ topology, supporting
earlier considerations~\cite{Cheng:2009yz,Belle:2007lyc}. 
The nearly identical branching fractions, as shown in Eq.~(\ref{data1}), 
also provide consistent evidence.
In our latest findings, we unveil additional insights. For $\bar B^0_s\to \Xi_c^{0(+)}\bar \Xi_c^{0(-)}$, 
the interference of the $W_{\rm ex}$ amplitude with the $W_{\rm em}$ amplitude 
adds a new contribution to the decay process. 
Furthermore, $\bar B^0_s\to \Lambda_c^+\bar \Lambda_c^-$ 
represents a pure $W_{\rm ex}$ decay, 
offering a clear and distinct case for experiments to clarify 
if the $W_{\rm ex}$ contribution can be neglected.

The $b\to c\bar c d$ induced decays with $|V_{cd}/V_{cs}|\simeq 0.05$ are more suppressed.
Unlike ${\cal M}(\bar B^0_s\to \Xi_c^0\bar \Xi_c^0)={\cal M}(\bar B^0_s\to \Xi_c^+\bar \Xi_c^-)$,
$\bar B^0\to \Xi_c^+\bar \Xi_c^-$ as a pure $W_{\rm ex}$ decay
does not share an isospin relation with $\bar B^0\to \Xi_c^0\bar \Xi_c^0$,
which is against our naive expectation. Instead, the equality relation arises from 
${\cal M}(\bar B^0\to \Lambda_c^+\bar \Lambda_c^-)={\cal M}(\bar B^0\to \Xi_c^0\bar \Xi_c^0)$.
Moreover, an interesting triangle relation exists:
\begin{eqnarray}\label{TR}
{\cal M}(\bar B^0\to \Xi_c^+\bar \Xi_c^-)+{\cal M}(B^-\to \Xi_c^0\bar \Xi_c^-)
={\cal M}(\bar B^0\to \Xi_c^0\bar \Xi_c^0)\,.
\end{eqnarray}
If the $W_{\rm{ex}}$ contribution is negligible, 
the relation is simplified to ${\cal M}(B^-\to \Xi_c^0\bar \Xi_c^-)
\simeq {\cal M}(\bar B^0\to \Xi_c^0\bar \Xi_c^0)$, resulting in nearly equal branching fractions.

It can be challenging to compute the $W_{\rm ex}$ and $W_{\rm em}$ amplitudes. 
For example, the factorization approach derives the $W_{\rm ex}$ amplitude as
${\cal M}_{\rm wex}\propto f_B q^\mu\langle {\bf B}_c
\bar{\bf B}'_c|\bar c\gamma_\mu(1-\gamma_5)c|0\rangle$~\footnote{Please consult 
the similar deviation for ${\cal M}(B\to{\bf B}_c \bar{\bf B})$ in Ref.~\cite{Hsiao:2019wyd}.},
where $f_B$ is the B meson decay constant, $q^\mu$ the momentum transfer, and
the matrix elements present the vacuum (0) to ${\bf B}_c\bar{\bf B}'_c$ production.
As information on the $0\to {\bf B}_c\bar{\bf B}'_c$ production is lacking, 
a model calculation is currently unavailable. For a calculation on ${\cal M}_{\rm wem}$,
one proposes a meson propagator to provide an additional quark pair, 
resulting in the branching fractions to be a few times $10^{-3}$  
for $\bar B^0\to \Xi_c^+\bar \Lambda_c^-$ and $B^-\to \Xi_c^0\bar \Lambda_c^-$~\cite{Cheng:2009yz}.
Additionally, a theoretical attempt incorporating final state interactions yields  
${\cal B}\simeq {\cal O}(10^{-3})$~\cite{Chen:2006fsa}.
It appears that the above approaches may overestimate the $W_{\rm em}$ contribution.

Without relying on the aforementioned model calculations, 
we determine $e$ and $c'$ using the experimental data based on the $SU(3)_f$ symmetry,
as described in Eq.~(\ref{para2}) and the surrounding context. 
By employing $|c'|=(1.29\pm 0.18)~{\rm GeV}^3$, we successfully replicate
${\cal B}(\bar B^0\to \Xi_c^+\bar \Lambda_c^-)=(7.2^{+2.1}_{-1.9})\times 10^{-4}$ and 
${\cal B}(B^-\to \Xi_c^0\bar \Lambda_c^-)=(7.8^{+2.3}_{-2.0})\times 10^{-4}$,
consistent with the experimental inputs. This demonstration underscores that 
$c'$ can effectively estimate the single $W_{\rm em}$ contribution.
For further examination, we explore other decay channels 
that receive the single $W_{\rm em}$ contribution. 
We hence predict the following branching fractions:
\begin{eqnarray}
{\cal B}(B^-\to \Xi_c^0\bar \Xi_c^-)
&=&(3.4^{+1.0}_{-0.9})\times 10^{-5}\,, \nonumber\\
{\cal B}(\bar B^0_s\to \Lambda_c^+\bar \Xi_c^-)
&=&(3.9^{+1.2}_{-1.0})\times 10^{-5}\,,
\end{eqnarray}
which are promising to be measured by experimental facilities such as LHCb.

The previous studies have assumed that $\bar B^0\to \Lambda_c^+\bar \Lambda_c^-$
receives the single $W_{\rm em}$ contribution~\cite{Cheng:2009yz,Belle:2007lyc}. 
Consequently, the estimated branching fraction
${\cal B}(\bar B^0\to \Lambda_c^+\bar \Lambda_c^-)\simeq 5\times 10^{-5}$
mentioned in the introduction significantly exceeds the experimental upper bound.
In Table~\ref{tab1}, since ${\cal M}(\bar B^0\to \Lambda_c^+\bar \Lambda_c^-)
=-\lambda_{cd} (2e+c')$ is found to include the $SU(3)_f$ parameter~$e$, 
it suggests a non-negligible $W_{\rm ex}$ amplitude. By newly incorporating $e$, 
a destructive interference with $c'$ could occur, effectively reducing 
the branching fraction. In fact, we estimate $|e|\simeq 0.2$~GeV$^3$, 
and obtain
\begin{eqnarray} 
{\cal B}(\bar B^0\to \Lambda_c^+\bar \Lambda_c^-)
=(2.1^{+1.0}_{-0.8})\times 10^{-5}\,, 
\end{eqnarray}
thus alleviating the discrepancy. 

To carefully test the $W_{\rm ex}$ contribution, we predict
the branching fractions of the other interfering decay channels:
\begin{eqnarray}
&&{\cal B}(\bar B^0_s\to \Xi_c^{0(+)}\bar \Xi_c^{0(+)})
=(3.0^{+1.4}_{-1.1})\times 10^{-4}\,,\nonumber\\
&&{\cal B}(\bar B^0\to \Xi_c^0\bar \Xi_c^0)
=(1.5^{+0.7}_{-0.6})\times 10^{-5}\,.
\end{eqnarray}
When $|e|=0$, ${\cal B}(\bar B^0_s\to \Xi_c^{0(+)}\bar \Xi_c^{0(+)})$
would be enhanced to $(6.3^{+1.9}_{-1.6})\times 10^{-4}$; moreover,
${\cal B}(\bar B^0\to \Xi_c^0\bar \Xi_c^0)$ would be enhanced to
$(3.1^{+0.9}_{-0.8})\times 10^{-5}$, making it close to ${\cal B}(B^-\to \Xi_c^0\bar \Xi_c^-)$,
in accordance with the description for the triangle relation in Eq.~(\ref{TR}). 
We also anticipate non-zero branching fractions of the pure $W_{\rm ex}$ decays, 
given by
\begin{eqnarray}
&&
{\cal B}(\bar B^0_s\to \Lambda_c^+\bar \Lambda_c^-)
=(8.1^{+1.7}_{-1.5})\times 10^{-5}\,,\nonumber\\
&&
{\cal B}(\bar B^0\to \Xi_c^+\bar \Xi_c^-)=(3.0\pm 0.6)\times 10^{-6}\,,
\end{eqnarray}
which serve to test the $W$-exchange mechanism
in the $B\to{\bf B}_c\bar{\bf B}'_c$ decays.

To initiate a theoretical investigation of baryonic $B_c^+$ decays, 
we derive the amplitudes for $B_c\to {\bf B}_c\bar{\bf B}'_c$
using the $SU(3)_f$ symmetry. This results in two possible decay channels:
$B^+_c\to \Xi_c^+\bar \Xi_c^0$ and $B^+_c\to \Lambda_c^+\bar \Xi_c^0$,
with $\bar c'$ representing the sole contribution from the $W_{\rm em}$ term, as given in Table~\ref{tab1}.
Upon comparing the topologies in Fig.~\ref{fig1}b and Fig.~\ref{fig1}c, 
an evident similarity emerges between 
$B_c\to{\bf B}_c\bar{\bf B}'_c$ and $B\to{\bf B}_c\bar{\bf B}'_c$. 
Both decays involve $c\bar c$ in the final states, 
where the charm quark pair is not part of the $SU(3)_f$ symmetry.
Additionally, both decays necessitate extra gluons connecting 
$q\bar q$ in the ${\bf B}_c\bar {\bf B}'_c$ formations, 
suggesting similar QCD effects during the hadronization processes.
Thus, while $\bar c'$ and $c'$ in Eq.~(\ref{amp}) are derived as two seemingly different parameters 
within the $SU(3)_f$ symmetry, it is reasonable to assume that $\bar c' = c'$, 
supported by the resemblance between the topologies in Fig.~\ref{fig1}b and Fig.~\ref{fig1}c.
With $\bar c' = c'$, we predict the following branching fractions:
\begin{eqnarray}
{\cal B}(B^+_c\to \Xi_c^+\bar \Xi_c^0) &=& (2.8^{+0.9}_{-0.7})\times 10^{-4}\,, \nonumber\\
{\cal B}(B^+_c\to \Lambda_c^+\bar \Xi_c^0) &=& (1.6^{+0.5}_{-0.4})\times 10^{-5}\,.
\end{eqnarray}
These predictions can be investigated by the LHCb experiment.

\newpage
In summary, we have explored the two-body doubly charmful baryonic $B\to{\bf B}_c\bar{\bf B}'_c$ decays. 
Here, the $W_{\rm ex}$ and $W_{\rm em}$ amplitudes have been parametrized as $e$ and $c'$, 
respectively, using the $SU(3)_f$ approach. With the determination of the $SU(3)_f$ parameters, 
we have calculated the branching fractions 
${\cal B}(B^-\to \Xi_c^0\bar \Xi_c^-)=(3.4^{+1.0}_{-0.9})\times 10^{-5}$ and 
${\cal B}(\bar B^0_s\to \Lambda_c^+\bar \Xi_c^-)=(3.9^{+1.2}_{-1.0})\times 10^{-5}$. 
Considering that the single $W_{\rm em}$ contribution to $\bar B^0\to\Lambda_c^+\bar\Lambda_c^-$ 
has caused its branching fraction to significantly exceed the experimental upper bound, 
we have added the $W_{\rm ex}$ amplitude (or the $SU(3)_f$ parameter $e$), 
overlooked in previous studies, 
to destructively interfere with the $W_{\rm em}$ amplitude.
As a consequence, we have alleviated the discrepancy. 
To further test the interfering decay channels, 
we have predicted ${\cal B}(\bar B^0_s\to \Xi_c^{0(+)}\bar \Xi_c^{0(+)})
=(3.0^{+1.4}_{-1.1})\times 10^{-4}$ and 
${\cal B}(\bar B^0\to \Xi_c^0\bar \Xi_c^0)
=(1.5^{+0.7}_{-0.6})\times 10^{-5}$. 
For the pure $W_{\rm ex}$ decay channels, 
we have expected non-zero branching fractions, such as 
${\cal B}(\bar B^0_s\to \Lambda_c^+\bar \Lambda_c^-)=(8.1^{+1.7}_{-1.5})\times 10^{-5}$ and 
${\cal B}(\bar B^0\to \Xi_c^+\bar \Xi_c^-)=(3.0\pm 0.6)\times 10^{-6}$, 
promising to be observed in near-future measurements. 
Additionally, we have predicted 
${\cal B}(B^+_c\to \Xi_c^+\bar \Xi_c^0)=(2.8^{+0.9}_{-0.7})\times 10^{-4}$ and 
${\cal B}(B^+_c\to \Lambda_c^+\bar \Xi_c^0)=(1.6^{+0.5}_{-0.4})\times 10^{-5}$, 
which are accessible to the LHCb experiment.

\section*{ACKNOWLEDGMENTS}
This work was supported by NSFC (Grants No.~11675030 and No.~12175128).

\end{document}